\documentclass[letterpaper]{article} % DO NOT CHANGE THIS
\usepackage[]{aaai2026}  % DO NOT CHANGE THIS
\usepackage{times}  % DO NOT CHANGE THIS
\usepackage{helvet}  % DO NOT CHANGE THIS
\usepackage{courier}  % DO NOT CHANGE THIS
\usepackage[hyphens]{url}  % DO NOT CHANGE THIS
\usepackage{graphicx} % DO NOT CHANGE THIS
\usepackage{natbib}  % DO NOT CHANGE THIS AND DO NOT ADD ANY OPTIONS TO IT
\usepackage{caption} % DO NOT CHANGE THIS AND DO NOT ADD ANY OPTIONS TO IT
\usepackage{algorithm}
\usepackage{algorithmic}
\usepackage{booktabs}
\usepackage{multirow}
\usepackage{amsmath, amssymb}
\usepackage{newfloat}
\usepackage{listings}
\DeclareCaptionStyle{ruled}{labelfont=normalfont,labelsep=colon,strut=off} % DO NOT CHANGE THIS
\floatstyle{ruled}
\newfloat{listing}{tb}{lst}{}
\floatname{listing}{Listing}
\nocopyright
\title{No Attention, No Problem: Rethinking Session-based Recommendation\\ with Pure Convolution}
\author {
    Tao Huang\textsuperscript{\rm 1},
    Wei Zhou\textsuperscript{\rm 1},
}
\affiliations {
    \textsuperscript{\rm 1}School of Big data and Software Engineering, Chongqing University, Chongqing, China\\
    htao@stu.cqu.edu.cn, zhouwei@cqu.edu.cn
}
\begin{document}

\maketitle

\begin{abstract}
Session-based recommendation (SBR) predicts the next choice in a session by analyzing recent interactions. Transformer-based models are widely used because of their ability to capture long-range dependencies through self-attention mechanisms. In contrast, traditional convolutional models, although more efficient, are often limited by their weak global modeling capabilities and are losing ground in SBR tasks. In this work, we propose a \textbf{Next}-generation Pure \textbf{Conv}olutional Framework (NextConvRec) for SBR tasks, aiming to balance efficiency and performance. NextConvRec uses a \textbf{S}tructural and \textbf{P}ositional \textbf{C}onvolutional \textbf{E}ncoder (SPCE) for preprocessing, combining learnable convolutional positional biases with session-level structural signals extracted through GCN layers. Its backbone convolutional module effectively expands the effective receptive field through depthwise convolutions and pointwise convolutions, enabling robust long-range preference modeling without attention mechanisms. Extensive experiments on 4 benchmark datasets show that NextConvRec outperforms several state-of-the-art baselines by around 1.73\% on average, and reduces the average inference time per session by 16.7\%. The convolutional architectures remain a promising direction for efficient and accurate session-based recommendations.
\end{abstract}


\section{Introduction}
Session-based Recommendation aims to provide personalized recommendations based on users' limited interactions within a short period in anonymous and temporary session scenarios. It has attracted widespread attention due to its significant practical value \cite{li2017neural, hidasi2015session, wang2022sequential, jannach2017session, gao2023survey}. It has a wide range of applications, such as e-commerce, media, entertainment, tourism, and finance \cite{wang2022sequential}. The emergence of models such as SASRec has led to Transformer-based and MLP-based models dominating the SBR task \cite{kang2018self, sun2019bert4rec, wu2020deja, zhou2022filter, fan2022sequential, du2023frequency, shin2024attentive, feng2024rotan, jiang2024trimlp}, while convolution-based models have achieved limited success in SBR tasks.

\begin{figure}[htbp]
\centering
\includegraphics[width=0.48\textwidth]{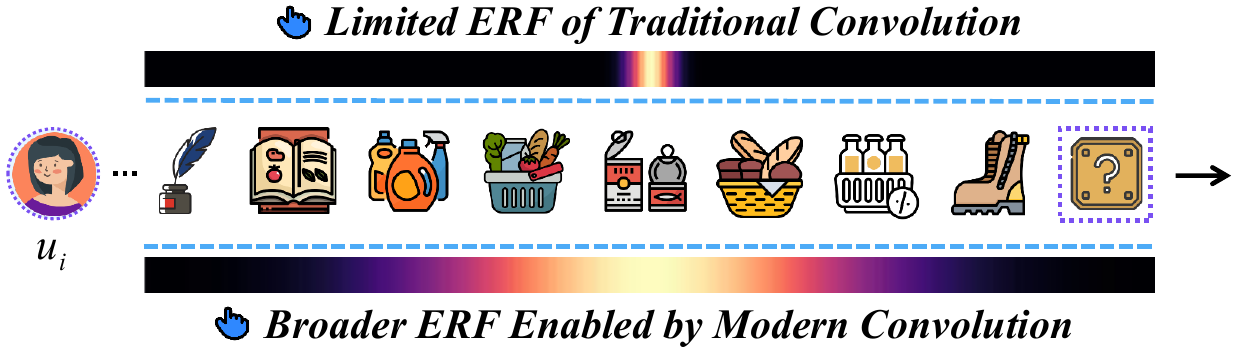} % Reduce the figure size so that it is slightly narrower than the column.
\caption{\small An explanation of the application of convolution technology in SBR tasks. Modern convolution technology has a larger Effective Receptive Field (ERF), enabling better context modeling.}
\label{fig:intro}
\end{figure}

Convolution is a pioneering neural structure that was used early on in SBR tasks \cite{tang2018personalized}. It is crucial for session-based recommendations because it better balances efficiency and performance, effectively learning local patterns to achieve efficient and robust recommendations. Convolutional models and their variants were widely adopted in SBR tasks in the early 2010s. However, the landscape changed with the emergence of Transformer-based models such as SASRec. Transformer-based models feature a global effective receptive field, enabling them to better capture global session patterns and achieve remarkable performance, significantly outperforming traditional convolutional models. As a result, traditional convolutional models have gradually lost their prominence in SBR tasks.

Dating back to the 2010s, CNNs have many well-known applications in SR tasks. Inspired by the successful application of CNNs in image tasks, the representative work Caser abandoned the RNN structure and proposed a convolutional sequence embedding model, proving that CNN-based recommendation models can achieve superior performance \cite{tang2018personalized}. NextItNet is an extension of Caser, employing deep one-dimensional convolutional stacks combined with dilated convolutions to effectively expand the receptive field while maintaining efficient parallel computation \cite{yuan2019simple}. Additionally, there are a few other works, such as 3D-CNN, that attempt to leverage CNNs to process auxiliary information \cite{tuan20173d}. From these works, it is evident that the application of CNNs in SBR remains limited. 

From the perspective of computer vision (CV), this is a field where convolutional technology is widely used and actively updated. Unlike the application of convolutions in the field of session recommendations, the field of computer vision focuses on optimizing convolutions themselves and has proposed modern convolution techniques \cite{liu2022convnet,liu2022more,ibtehaz2023acc}. Modern convolutions represent a new convolutional paradigm inspired by Transformer-based models, featuring two primary improvements: 
(1) adopting certain design elements from the Transformer architecture, replacing multi-head attention blocks with modern convolutional blocks; and 
(2) employing a large convolutional kernel to effectively increase the effective receptive field. Modern convolution techniques have also been widely transferred and applied to other fields, such as time series prediction \cite{luo2024moderntcn,cheng2025convtimenet} and Generative Adversarial Networks (GANs) \cite{huang2024gan}. The effectiveness of modern convolution has been widely validated.

According to the bucket principle, to bring convolutions back to the SBR stage, it is necessary to retain their efficient parallel computing and local pattern modeling advantages while addressing the shortcomings of their limited effective receptive field (ERF) (as shown in Fig.~\ref{fig:intro}). Based on the research overview, introducing and improving modern convolution techniques can achieve a better balance between performance and efficiency. The value of convolutional neural networks in SBR is not to surpass Transformers, but to achieve a solution that matches their performance while significantly outperforming them in inference efficiency. Therefore, our goal is to construct a convolutional model for SBR tasks that balances performance with efficient inference.

Based on the above motivations, we designed a pure convolutional framework called NextConvRec for SBR tasks. Specifically, we adapted modern convolutional techniques to traditional convolutional models and made some modifications based on the characteristics of the SBR domain to leverage its potential better. Additionally, we have designed a specialized convolutional preprocessing structure, the structural and positional convolutional encoder (SPCE), for modern convolutional modules, enabling modern convolutional structures to better capture long-range preferences.

To comprehensively evaluate the effectiveness of NextConvRec, we conducted large-scale experiments on 4 benchmark datasets. The results show that NextConvRec can rival the latest Transformer-based state-of-the-art (SOTA) models in terms of recommendation performance while significantly outperforming existing Transformer-based models in terms of inference efficiency. Additionally, we demonstrate the necessity of the proposed method through a series of experiments and delve into several key hyperparameters of convolutional models. Furthermore, we compare and analyze the inference speed and convergence efficiency of different models. In summary, the main contributions of this paper are as follows:
\begin{itemize}
    \item We revisit the application of modern convolutions in session-based recommendation tasks and propose a pure convolutional model, NextConvRec, which fully exploits and unleashes the potential of convolutional structures.
    \item We designed a convolutional encoder (SPCE) that combines graph structure and positional awareness as a preprocessing layer, effectively enhancing the compatibility and expressiveness of convolutional and session graph modeling.
    \item Extensive experiments on 4 real-world datasets demonstrate that NextConvRec maintains the efficiency advantages of convolutional models, significantly improves inference efficiency, and achieves performance comparable to state-of-the-art Transformer models, thereby achieving a better balance between performance and efficiency.
\end{itemize}

\section{Related Work}
\subsection{Session-based Recommendation}
Due to strict privacy policies, Session-Based Recommendations were proposed to address the difficulty of accessing sensitive identity information. SBR can predict the next item of interest for an anonymous user based on their limited behavior over a short period \cite{li2017neural,hidasi2015session,wang2021survey}. Early methods combined Markov chains to capture transitions between items \cite{rendle2010factorizing,he2016fusing}. Additionally, time is an important indicator of changes in user preferences and is emphasized in the SBR field \cite{li2017neural,hidasi2015session,dallmann2017improving}. Temporal models are categorized into implicit and explicit types. The former relies on RNNs and their variants, such as GRURec \cite{hidasi2015session}, NARM \cite{li2017neural}, and RepeatNet \cite{ren2019repeatnet}, while the latter explicitly injects positional embeddings into the original embeddings, such as models based on KNN \cite{garg2019sequence} and GNN \cite{chen2020handling,li2022spatiotemporal}. CNN is another excellent neural architecture used in the early stages of SBR, characterized by efficient parallelization. Notable examples include Caser \cite{tang2018personalized}, NextItNet \cite{yuan2019simple}, and 3D-CNN \cite{tuan20173d}. In recent years, some pure model-based approaches for SBR have been proposed, such as the pure MLP model FMLP4Rec \cite{zhou2022filter} and TriMLP \cite{jiang2024trimlp}. Additionally, large language models have been widely applied in the SBR field \cite{wang2025re2llm,liu2025llmemb,ye2025harnessing}, as well as diffusion models \cite{ma2024plug}.

\subsection{Transformer-based Method}
With the emergence of Transformer-based models such as SASRec \cite{kang2018self} and BERT4Rec \cite{sun2019bert4rec}, the effectiveness of self-attention has been proven. Transformer-based models outperform convolution-based models due to their superior global modeling capabilities, establishing a new paradigm \cite{liu2021noninvasive,zhang2023beyond}. Transformer-based models have many excellent variants, such as DSAN \cite{yuan2021dual}, which introduces a dual sparse attention mechanism; FEARec \cite{du2023frequency} and BSARec \cite{shin2024attentive}, which enhance performance through filters; and CSRec \cite{liu2025csrec}, which introduces causal sequence recommendation. IFCDSR \cite{wu2025image} combines item image information.

\subsection{Modern Convolution}
Modern convolutional techniques were first proposed in the field of computer vision \cite{liu2022convnet,liu2022more,ibtehaz2023acc}, significantly improving their modeling capabilities through modifications to the convolutional structure. In recent years, modern convolution techniques have been widely migrated and applied to multiple fields, e.g., DCNv4 \cite{xiong2024efficient}, CFSR \cite{wu2024transforming}, and PeLK \cite{chen2024pelk} in computer vision, ModernTCN \cite{luo2024moderntcn} and ConvTimeNet \cite{cheng2025convtimenet} in time series prediction, and Generative Adversarial Networks (GANs) \cite{huang2024gan}, among others. However, in the field of session-based recommendation, convolutional models have long lacked updates and attention, and this study aims to address this research gap.

\section{Preliminaries}
\subsection{Problem Definition}
In Session-Based Recommendation tasks, the objective is to predict the next item that a user is most likely to interact with, given only the sequence of interactions within the current session.  Let $\mathcal{V}$ denote the set of all items, and let the following expression represent the ordered sequence of items interacted with in a single session: $S^u = \{ v^u_1, v^u_2, \dots, v^u_{|S^u|} \}$, where $v^u_i \in \mathcal{V}$ denotes the $i$-th interacted item in the session and $|S^u|$ is the session length.  
The goal of SBR is to recommend the most probable next item $v \in \mathcal{V}$ at step $|S^u| + 1$ based on the session context $S^u$, which can be formulated as: $\hat{v} = \arg\max_{v \in \mathcal{V}} P(v \mid S^u)$. In practical recommendation scenarios, the model produces a ranked list of candidate items, and the top-$k$ items are presented to the user as potential next interactions.

\subsection{Modeling Session and Cross-session Graphs}
\noindent{\textbf{Session Graph $G_{s}$}}\quad For SBR tasks, we typically convert a sequence of sessions into a directed graph structure $G_{s}$. The nodes of this graph are uniquely identified items in the session, and the direction of the edges represents the click order of adjacent items. The edge weight $w_{ij}$ from node $i$ to node $j$ is defined as: $w_{ij} = \frac{\text{count}(i \rightarrow j)}{\text{outdeg}(i)}$, where $\mathrm{count}(i \rightarrow j)$ represents the number of transitions from $i$ to $j$ in the session, and $\mathrm{outdeg}(i)$ represents the out-degree of node $i$. Finally, we concatenate the in-edge weight matrix $A_{\mathrm{in}} \in \mathbb{R}^{n \times n}$ and the out-edge weight matrix $A_{\mathrm{out}} \in \mathbb{R}^{n \times n}$ to obtain the adjacency matrix for the session ($n$ denotes the number of unique items appearing in the session).

\noindent{\textbf{Relationship Graph $G_{r}$}}\quad Relationship Graph $G_r$ is used to describe the semantic relevance between different sessions. If there is at least one item that is common to two different sessions, an undirected edge is established between the two session nodes. The higher the edge weight, the closer the potential interest patterns between the sessions.

\subsection{Convolution for Session-based Recommendation}
In the SBR scenario, a session is usually represented as: $\text{X}\in {{\mathbb{R}}^{L\times D}}$, where $L$ represents the session length and $D$ represents the feature dimension of each position, also known as the channel. The convolution operation slides a local convolution kernel along the sequence dimension, encoding adjacent interactions. We can generally represent the CNN processing process as follows:
\begin{equation}
    \mathbf{o} = \mathrm{Conv}([\mathbf{v}_1; \mathbf{v}_2; \dots; \mathbf{v}_t]),
\end{equation}
where $[;]$ denotes the concatenation operation, $\mathrm{Conv}(\cdot)$ is the convolutional layer including stacked filters and pooling operations, and $\mathbf{o}$ is the output, which is expected to encapsulate the local patterns of user actions. To enhance modeling capabilities, convolutional modules typically use depthwise separable convolution: i) Depthwise convolution performs one-dimensional convolution independently on each channel to capture time dependencies within a single feature dimension:
\begin{equation}
    \mathbf{Z}^{(dw)} = \mathrm{DWConv}(\mathbf{X}, \ \mathrm{groups}=d),
\end{equation}
ii) Pointwise convolution uses 1×1 convolution to fuse information from different channels:
\begin{equation}
    \mathbf{Z}^{(pw)} = \mathrm{PWConv}(\mathbf{Z}^{(dw)}).
\end{equation}
When $n$ layers of convolutional modules are stacked, the effective receptive field is approximately: $\mathrm{ERF} \approx 1 + n \cdot (k-1)$. By increasing the number of stacked layers, longer session dependencies can be gradually captured while maintaining low computational complexity.

\section{Proposed Method}
\begin{figure}[htbp]
\centering
\includegraphics[width=0.48\textwidth]{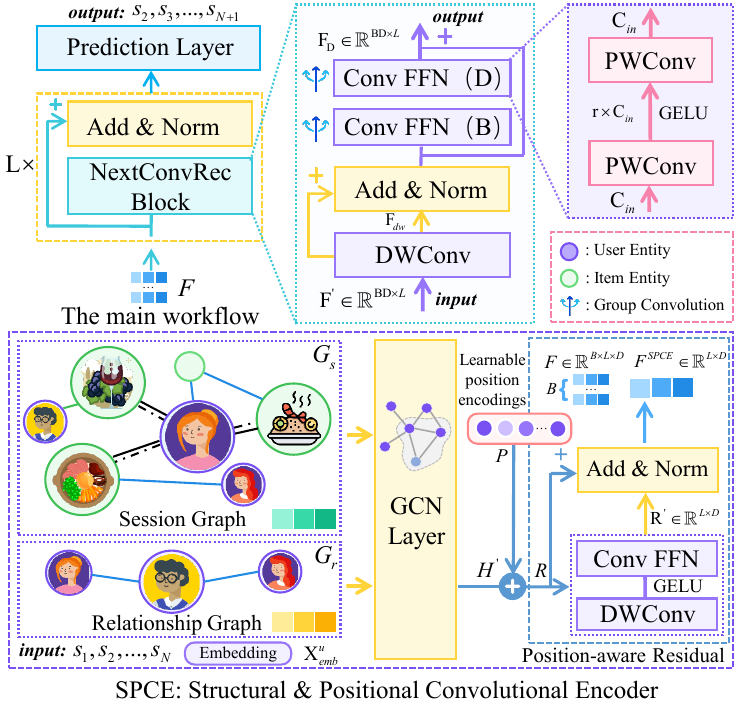} % Reduce the figure size so that it is slightly narrower than the column.
\caption{\small The framework design of NextConvRec that we proposed. NextConvRec is a pure convolutional session-based recommendation framework, mainly consisting of an embedding preprocessing layer SPCE and a backbone convolutional module NextConvRec Block. Conv FFN is divided into B group and D group, which learn dependencies between channels and sessions, respectively.}
\label{fig:main_framework}
\end{figure}

This section presents an overview of the NextCovRec framework (as shown in Fig.~\ref{fig:main_framework}) and the detailed design of the Structural \& Positional Convolution Encoder and NextConvRec Block.

\subsection{Embedding Layer}
Given a user's session interaction sequence $S^u=[v^u_1, v^u_2, \dots, v^u_{|S^u|}]$, considering that the historical session lengths of different users are inconsistent, we set the maximum session length to $L$, and if the session length is insufficient, we fill it with 0. Using the item embedding matrix $\mathbf{E} \in \mathbb{R}^{|\mathcal{V}| \times D}$, we denote the user's session as $X_{u}$, where $\text{X}_{i}^{\text{u}}={{E}_{{{\text{v}}_{i}}}}$. The embedding layer can be represented by the following equation:
\begin{equation}
    Embed({{\text{X}}^{u}})\to X_{emb}^{u}\in {{\mathbb{R}}^{L\times D}},
\end{equation}
where $\text{X}_{emb}^{u}$ represents the session representation matrix, $D$ denotes embedding dimension, and the filled action is encoded using a series of zero vectors and excluded in the gradient update step.

\subsection{Structural and Positional Convolutional Encoder}
The Structural and Positional Convolutional Encoder (SPCE) preprocesses embeddings based on the embedding layer and consists of two modules: a graph convolutional layer (GCN) and a position-aware residual. The GCN layer extracts higher-level user-item co-occurrence patterns based on session graphs \cite{wu2019session} and relationship graphs \cite{liu2021item}.

\noindent{\textbf{Construction of Two Types of Graphs}}\quad Based on the definitions in preliminaries, we construct two graphs needed for the model. First, for each session, we generate a directed graph session graph $G_{s}$ based on the transition relationship between adjacent items. Second, we construct a relationship graph $G_r$, representing each session as a node and connecting two sessions that share at least one common item. The edge weight is defined as the ratio of the number of co-occurring items to the total number of items in the two sessions.

\noindent{\textbf{GCN Layer}}\quad We introduce a graph convolution layer based on Gated Graph Neural Network (GGNN) \cite{li2015gated}, defined as follows:
\begin{align}
    \mathbf{X} &= [\mathbf{A} \mathbf{H} \,\|\, \mathbf{H}], \\
    \mathbf{r} &= \sigma(\mathbf{W}_r \mathbf{X} + \mathbf{U}_r \mathbf{H}), \\
    \mathbf{z} &= \sigma(\mathbf{W}_z \mathbf{X} + \mathbf{U}_z \mathbf{H}), \\
    \tilde{\mathbf{H}} &= \tanh(\mathbf{W}_n \mathbf{X} + \mathbf{U}_n (\mathbf{r} \odot \mathbf{H})), \\
    \mathbf{H}' &= (1 - \mathbf{z}) \odot \mathbf{H} + \mathbf{z} \odot \tilde{\mathbf{H}},
\end{align}
where $\mathbf{H} \in \mathbb{R}^{L \times D}$ is the initial item embedding of the session, $\mathbf{A} \in \mathbb{R}^{L \times L}$ is the adjacency matrix, $L$ denotes the session length, $\sigma(\cdot)$ is the Sigmoid function, $\odot$ denotes element-wise multiplication, $\mathbf{W}_*$ and $\mathbf{U}_*$ are learnable parameters, and $r$ and $z$ are the reset gate and update gate, respectively.

\noindent{\textbf{Position-aware Residual}}\quad Standard position embedding is static and redundant for convolutional structures (DWConv already implicitly captures sequence offset awareness), so we assign learnable position encodings $\mathbf{P}$ to each position in the structure representation $\mathbf{H}'$, and then capture local dependency patterns between different positions through depthwise convolutions (Depthwise Conv) and GELU \cite{hendrycks2016gaussian} as follows:
\begin{align}
\mathbf{R} &= \mathbf{H}' + \mathbf{P},\\
\mathbf{R}' &= \text{ConvFFN}(\text{GELU}(\text{DWConv}(\mathbf{R}))),
\end{align}
Then, residual connections and layer normalization are used to stabilize training and retain original information:
\begin{equation}
\mathbf{F}^{SPCE} = \text{LayerNorm}(\mathbf{R} + \text{Dropout}(\mathbf{R}')),
\end{equation}
where ${\mathbf{F}^{SPCE}}\in {{\mathbb{R}}^{L\times D}}$ is a session representation that combines structural information and positional awareness, which is used as input for the subsequent NextConvRec Block. $\mathbf{F}\in {{\mathbb{R}}^{B\times L\times \text{D}}}$denotes the batched feature representation, where $B$ is the batch size.

\subsection{Next-generation Convolutional Block}
NextCov Block is a pure convolutional structure designed to balance efficiency and performance. Based on the idea of modern convolutions  \cite{liu2022convnet, luo2024moderntcn}, we replace the multi-head attention block of Transformer-based models in session-based recommendation tasks with depth-separable convolutions (including depthwise convolutions and pointwise convolutions) to accomplish three tasks: (1)\ Learning temporal dependencies within each session, (2)\ learning dependencies between dimensions, and (3)\ learning dependencies between sessions.

\noindent{\textbf{Depthwise Convolution}}\quad Depthwise Convolution is used to learn the temporal dependencies within each session. For a single session, we reshape the input $F\in {{\mathbb{R}}^{B\times L\times D}}$ to $F\in {{\mathbb{R}}^{BD\times L}}$, then set $group = BD$ to make features and variables independent, enabling independent learning of the temporal dependencies of each univariate session. Additionally, we set a large convolution kernel ${{K}_{l}}$ to expand the effective receptive field (ERF). We perform $\mathbf{1D}$ convolution independently on each channel as follows:
\begin{equation}
    {\mathbf{F}_{dw}}=\text{DWConv}_{BD}(\mathbf{F},\text{Kernel}={\mathbf{K}_{l}}),
\end{equation}
where $\mathrm{DWConv}(\cdot)$ represents a depthwise convolution operation with the number of groups equal to the number of channels, ensuring that the convolution kernel of each channel only acts on the corresponding session.

\noindent{\textbf{ConvFFN}}\quad ConvFFN should complementarily mix information across feature and variable dimensions, so it should be decoupled into ConvFFN(B) and ConvFFN(D), with the former responsible for learning the feature representation of each action in the session and the latter responsible for capturing the cross-variable dependencies of each action in the session. Given the output $\mathbf{F}_{dw}$ of DWConv, a ConvFFN layer consisting of two consecutive pointwise convolutions (PWConv) is applied:
\begin{equation}
    {{\mathbf{F}}^{B}}\in {{\mathbb{R}}^{BD\times L}}=\text{ConvFFN}({\mathbf{F}_{dw}},\text{group}=B),
\end{equation}
where $\mathbf{F}^{B}$ is the output of the ConvFFN(B) layer, which learns the dependencies of each channel D in each session. Next, continue through a ConvFFN layer:
\begin{equation}
    {{\mathbf{F}}^{D}}\in {{\mathbb{R}}^{BD\times L}}=\text{ConvFFN}({\mathbf{F}_{B}},\text{group}=D),
\end{equation}
where $\mathbf{F}^{D}$ is the output of the ConvFFN(D) layer, which learns the dependencies between B sessions. In the ConvFFN calculation process, the channel uses a calculation method of first increasing the dimension and then decreasing it. $\mathbf{ffn}_{ratio}$ is a control hyperparameter of the magnification factor, which determines the size of the hidden dimension in the middle of the feedforward layer.

\subsection{Overall Structure}
Overall, the embedding $\mathbf{X}_{emb}^{u}$ learns dependencies across time, variables, and channels through the SPCE layer and NextCovRec block, ultimately yielding the informative representation ${{\mathbf{F}}^{D}}\in {{\mathbb{R}}^{BD\times L}}$. We reshape it to obtain the final informative representation $\mathbf{Y}\in {{\mathbb{R}}^{B\times L\times D}}$, and the entire process can be represented as follows:
\begin{equation}
    \mathbf{Y}=\text{NextConvRec}(\mathbf{X}_{emb}^{u})
\end{equation}
where $\mathrm{NextConv}(\cdot)$ denotes a stacked NextConvRec block. Each NextConvRec block is organized using residual connections \cite{he2016deep}, and the i-th NextConvRec block is represented as follows:
\begin{equation}
{{\mathbf{Y}}^{i+1}}=\text{NextConvRec}({\mathbf{Y}^{i}})+{\mathbf{Y}^{i}}
\end{equation}
where the final number of stacked layers is set to $K$.

\subsection{Prediction Layer}
In the last layer $K$ of NextConvRec, we calculate the item preference score from the user's historical sessions. The score is calculated using the following formula:
\begin{equation}
    \hat{y}_v = p\left(v_{|\mathcal{S}^u|+1}^u = v \mid \mathcal{S}^u\right) 
    = \mathbf{e}_v^\top \mathbf{Y}^{K}_{|\mathcal{S}^u|},
\end{equation}
where the dot product is used to measure the similarity between the project embedding $\mathbf{e}_v$ and the user's final representation $\mathbf{Y}^{\mathrm{K}}_{|\mathcal{S}^u|}$,
 thereby obtaining the preference score $\hat{y}_v$. For training, we use the CE loss function to optimize the model parameters \cite{du2023frequency, shin2024attentive}, as shown in the following formula:
\begin{equation}
    \mathcal{L} = -\log \frac{\exp(\hat{y}_g)}{\sum\limits_{i \in \mathcal{|V|}} \exp(\hat{y}_i)},
\end{equation}
 where $g \in |\mathcal{V}|$ is the ground-truth next item.

\section{Experiments and Analysis}
We conducted a series of experiments to prove the effectiveness of NextConvRec by exploring several problems. \textbf{RQ1}: Does the proposed model perform better than the baseline model in the session-based recommendation task? \textbf{RQ2}: How do various model enhancement methods or components affect the model's performance and effectiveness? \textbf{RQ3}: Can NextConvRec maintain stable performance across different receptive field sizes and feedforward capabilities? \textbf{RQ4}: Does the proposed model achieve faster inference speeds while maintaining accuracy, thereby providing a better balance between efficiency and performance? 

\subsection{Experimental Setup}
\noindent{\textbf{Datasets}}\quad We conduct experiments on four widely-used public session-based recommendation datasets: Amazon Beauty, Amazon Sports, Amazon Toys \cite{mcauley2015image}, and Yelp\footnote{https://www.yelp.com/dataset}. These datasets are commonly used in the SBR field to evaluate model performance at different levels of sparsity and domain diversity. The dataset statistics are shown in Table~\ref {tab:dataset_statistics}.

\begin{table}[htbp]
\centering
\caption{Statistics of datasets.}
\label{tab:dataset_statistics}
\small
\renewcommand{\arraystretch}{1.1}
\setlength{\tabcolsep}{4.5pt}  % 缩小列间距
\begin{tabular}{lrrrr}
\toprule
\textbf{Dataset} & \textbf{\#Users} & \textbf{\#Items} & \textbf{\#Actions} & \textbf{Avg. Len} \\
\midrule
Beauty             & 22364  & 12102  & 198502  & 8.9  \\
Sports \& Outdoors & 35598  & 18357  & 296337  & 8.3  \\
Toys \& Games      & 19413  & 11925  & 167597  & 8.63 \\
Yelp               & 30450  & 20039  & 316541  & 10.4 \\
\bottomrule
\end{tabular}
\end{table}

\noindent{\textbf{Baseline Methods}}\quad To validate the effectiveness of the model, we selected the following categories of state-of-the-art session-based recommendation models:
\begin{itemize}
    \item CNN/RNN-based models: \textbf{Caser}~\cite{tang2018personalized}, \textbf{GRU4Rec}~\cite{hidasi2015session}.
    \item Transformer-based models: \textbf{SASRec} \cite{kang2018self}, \textbf{BERT4Rec} \cite{sun2019bert4rec}, \textbf{FMLP4Rec} \cite{zhou2022filter}, \textbf{BSARec} \cite{shin2024attentive}.
    \item Intent-aware GNN-enhanced models: \textbf{ELCRec} \cite{liu2024end}.
\end{itemize}

\noindent{\textbf{Evaluation Metrics}}\quad We follow the earlier work in dividing the dataset \cite{kang2018self, shin2024attentive}, using the last item of the session for testing, the second-to-last item for validation, and the remaining items for training. To evaluate model performance, we use two widely adopted metrics: top-K Hit Rate (HR@K) and top-K Normalized Discounted Cumulative Gain (NDCG@K), where K = \{5, 20\}.

\noindent{\textbf{Implementation Details}}\quad Our model is implemented using PyTorch and trained on a Linux server equipped with three NVIDIA GeForce RTX 3090 GPUs (24 GB each). (1) In terms of basic parameter settings, we set the embedding dimension $D$ to 64 and the maximum session length $N$ to 50, truncating or padding any values below this threshold. (2) For hyperparameter settings, we stacked $L=2$ convolutional modules. The model adjusted small convolutional kernels and large convolutional kernels within the ranges \{3, 5, 7\} and \{15, 25, 31\}, respectively. The feed-forward expansion ratio (FFN ratio) was selected from \{1, 2, 4, 8\}. (3) In terms of inference efficiency, we record the average latency per session (in ms/session) at a fixed batch size. (4) For training, we use Adam optimization \cite{kingma2014adam}, with the learning rate selected from the set $\{5 \times 10^{-4}, 1 \times 10^{-3}\}$, and the batch size set to 256. 

\subsection{Overall Performance Analysis\ (RQ1)}

\begin{table*}[htbp]
\centering
\caption{Overall performance (HR@5/20 and NDCG@5/20) on four public datasets. The best result is in bold, and the best baseline is underlined. The \textit{Improv.} column reports the relative improvement (\%) of \textbf{NextConvRec} over the best baseline.}
\label{tab:main_results}
\small
\renewcommand{\arraystretch}{1.15}
\setlength{\tabcolsep}{3.3pt}
\resizebox{\textwidth}{!}{
\begin{tabular}{cccccccccccc}
\toprule
\multirow{2}{*}{\textbf{Dataset}} & \multirow{2}{*}{\textbf{Metric}} 
& \multicolumn{2}{c}{\textbf{CNN/RNN-based}} 
& \multicolumn{4}{c}{\textbf{Transformer-based\ \&\ MLP-based}} 
& \multicolumn{2}{c}{} & \multirow{2}{*}{\textbf{↑Improv.}} \\
\cmidrule(lr){3-4} \cmidrule(lr){5-8}
& & Caser & GRU4Rec & SASRec & BERT4Rec & FMLPRec & BSARec & ELCRec & \textbf{NextConvRec} \\
\midrule
\multirow{4}{*}{Beauty} 
& HR@5    & 0.0125 & 0.0169 & 0.0340 & 0.0469 & 0.0346 & \underline{0.0705} & 0.0529 & \textbf{0.0705} & +0.0\% \\
& HR@20   & 0.0403 & 0.0527 & 0.0823 & 0.1073 & 0.0869 & \underline{0.1314} & 0.1079 & \textbf{0.1324} & +7.6\% \\
& NDCG@5  & 0.0076 & 0.0104 & 0.0221 & 0.0311 & 0.0222 & \underline{0.0502} & 0.0355 & \textbf{0.0509} & +1.4\% \\
& NDCG@20 & 0.0153 & 0.0203 & 0.0356 & 0.0480 & 0.0369 & \underline{0.0673} & 0.0509 & \textbf{0.0683} & +1.5\% \\
\midrule
\multirow{4}{*}{Sports} 
& HR@5    & 0.0091 & 0.0118 & 0.0188 & 0.0275 & 0.0220 & \underline{0.0386} & 0.0286 & \textbf{0.0394} & +2.1\% \\
& HR@20   & 0.0260 & 0.0303 & 0.0459 & 0.0649 & 0.0525 & \underline{0.0801} & 0.0648 & \textbf{0.0789} & -1.5\% \\
& NDCG@5  & 0.0056 & 0.0079 & 0.0124 & 0.0180 & 0.0146 & \underline{0.0271} & 0.0185 & \textbf{0.0276} & +1.8\% \\
& NDCG@20 & 0.0104 & 0.0131 & 0.0200 & 0.0284 & 0.0231 & \underline{0.0387} & 0.0286 & \textbf{0.0397} & +2.8\% \\
\midrule
\multirow{4}{*}{Toys} 
& HR@5    & 0.0095 & 0.0121 & 0.0440 & 0.0412 & 0.0432 & \underline{0.076} & 0.0585 & \textbf{0.0779} & +2.5\% \\
& HR@20   & 0.0268 & 0.0348 & 0.0929 & 0.0939 & 0.0974 & \underline{0.1368} & 0.1138 & \textbf{0.1402} & +2.5\% \\
& NDCG@5  & 0.0058 & 0.0077 & 0.0297 & 0.0282 & 0.0288 & \underline{0.0553} & 0.0403 & \textbf{0.0565} & +2.2\% \\
& NDCG@20 & 0.0106 & 0.0140 & 0.0435 & 0.0430 & 0.0441 & \underline{0.0726} & 0.0560 & \textbf{0.0732} & +0.8\% \\
\midrule
\multirow{4}{*}{Yelp} 
& HR@5    & 0.0117 & 0.0130 & 0.0149 & 0.0256 & 0.0159 & \underline{0.0264} & 0.0236 & \textbf{0.0268} & +1.5\% \\
& HR@20   & 0.0337 & 0.0383 & 0.0424 & 0.0717 & 0.0490 & \underline{0.0729} & 0.0653 & \textbf{0.0738} & +1.2\% \\
& NDCG@5  & 0.0070 & 0.0080 & 0.0091 & 0.0159 & 0.0100 & \underline{0.0165} & 0.0150 & \textbf{0.0176} & +6.7\% \\
& NDCG@20 & 0.0131 & 0.0150 & 0.0167 & 0.0287 & 0.0192 & \underline{0.0295} & 0.0266 & \textbf{0.0299} & +1.4\% \\
\bottomrule
\end{tabular}
}
\end{table*}

Table~\ref{tab:main_results} shows the performance of NextConvRec and other baselines on four datasets. Based on the experimental results, we conducted the following analysis and discussion. 
(1) NextCovRec outperforms nearly all baseline models. Notably, NextCovRec surpasses the state-of-the-art Transformer-based baseline model BSARec in three out of four datasets. In the Yelp long-tail complex scenario, NextConvRec continues to maintain its advantage, with HR@20 and NDCG@5 improving by 1.2\% and 6.7\%, respectively. However, in the Sports dataset, NextConvRec's HR@20 metric is slightly lower than BSARec, possibly due to the longer session lengths in the Sports domain, which may be more suitable for attention-based modeling. Nevertheless, improvements are evident in other evaluation metrics. 
(2) As a classic CNN method, Caser performs poorly in terms of performance. Recently emerging Transformer-based and MLP-based models (such as BSARec and FMLPRec) have achieved significantly better performance than traditional convolutional models due to their global effective receptive fields (ERFs). BSARec is a significant advancement over SASRec, addressing the Transformer framework's limitations in extracting high-frequency signals. The NextCovRec model we propose represents an advancement of the classic CNN method in session-based recommendation. 
(3) Although ELCRec is a non-Transformer recommendation model that combines CNN and clustering mechanisms, its performance cannot match that of Transformer-based methods.

\subsection{Ablation Studies\ (RQ2)}
\begin{table}[htbp]
\centering
\caption{Ablation results(HR@20 and NDCG@20) of NextConvRec on three public datasets, where w/o indicates the removal of components and w/ indicates the replacement of components.}
\label{tab:ablation}
\small
\setlength{\tabcolsep}{3.3pt}
\renewcommand{\arraystretch}{1.15}
\begin{tabular}{lcccccc}
\toprule
\textbf{Method} & \multicolumn{2}{c}{\textbf{Beauty}} & \multicolumn{2}{c}{\textbf{Sports}} & \multicolumn{2}{c}{\textbf{Yelp}} \\
\cmidrule(lr){2-3} \cmidrule(lr){4-5} \cmidrule(lr){6-7}
& \textbf{H@20} & \textbf{N@20} & \textbf{H@20} & \textbf{N@20} & \textbf{H@20} & \textbf{N@20} \\
\midrule
\textit{w/o SPCE}       & 0.1303 & 0.0653 & 0.0746 & 0.0358 & 0.0709 & 0.0288 \\
\textit{w/o GCN}        & 0.1244 & 0.0629 & \underline{0.0752} & 0.0358 & 0.0688 & 0.0278 \\
\textit{w/o PosRes}     & 0.1280 & 0.0649 & 0.0769 & 0.0364 & 0.0713 & 0.0290 \\
\textit{w/ Trans} & 0.1262 & 0.0648 & 0.0745 & 0.0366 & 0.0700 & 0.0292 \\
\midrule
\textbf{Ours} & \textbf{0.1309} & \textbf{0.0672} & \textbf{0.0749} & \textbf{0.0368} & \textbf{0.0718} & \textbf{0.0293} \\
\bottomrule
\end{tabular}
\end{table}

We investigated the impact of each functional component on the overall performance of the model and defined the following four NextCovRec variant models: 
(1) \textbf{\textit{w/o SPCE}}: Removed the entire Structural and Position Convolutional Encoder (SPCE), leaving only the NextCovRec Block.
(2) \textbf{\textit{w/o GCN}}: Removed the GCN layer. 
(3) \textbf{\textit{w/o PosRes}}: Removing the Position-aware Residual processing of graph convolutional embeddings. 
(4) \textbf{\textit{w/ Trans}}: Removing all convolutional structures and replacing them with a multi-head attention mechanism structure. Table~\ref{tab:ablation} shows the performance comparison of NextCovRec and its variant models on the Beauty, Sports, and Yelp datasets. The results show that the preprocessing of the graph convolutional layer plays a crucial role. In addition, removing the learnable positional residual PosRes leads to a decrease in global sequence modeling ability (e.g., HR@20 decreased by 2.3\% in the Beauty dataset). Finally, replacing the modern convolutional backbone of NextConvRec with a standard multi-head attention module (MHSA) significantly reduces the model's performance and inference speed, indicating that attention mechanisms are not the optimal solution for balancing efficiency and performance in long-range modeling.

\subsection{Hyperparameter Sensitivity Analysis\ (RQ3)}
\begin{figure}[htbp]
\centering
\includegraphics[width=0.48\textwidth]{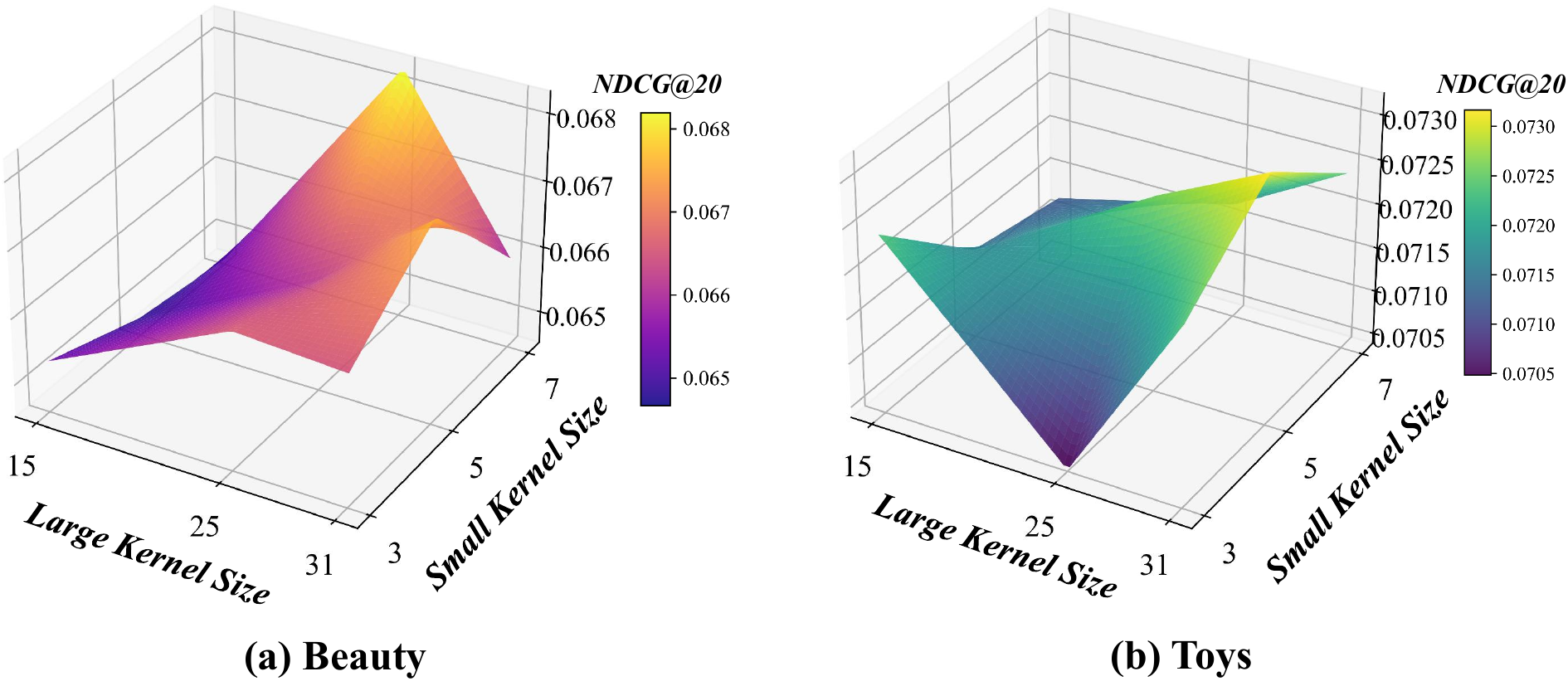} % Reduce the figure size so that it is slightly narrower than the column.
\caption{\small Sensitivity to small and large convolution kernel sizes. (NDCG@20 metric)}
\label{fig:Sensitivity-kernel_size}
\end{figure}

\noindent{\textbf{Convolution Kernel Size}}\quad We conducted a joint grid search for Small Kernel Size and Large Kernel Size. Fig.~\ref{fig:Sensitivity-kernel_size} shows the variation in NDCG@20 after varying the small kernel size and large kernel size. For Beauty, the optimal values for small and large kernel sizes are 7 and 25, respectively. For Toys, the optimal values for small and large kernel sizes are 5 and 31, respectively. Different datasets require controlling the two kernel sizes based on their characteristics to ensure complementarity between local feature extraction and global receptive field control. Increasing the large kernel size can effectively compensate for NextCovRec's global modeling capabilities.

\begin{figure}[htbp]
\centering
\includegraphics[width=0.48\textwidth]{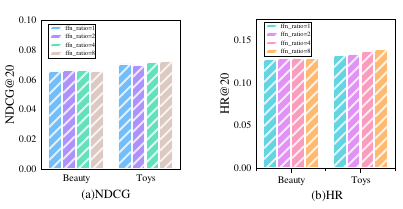} % Reduce the figure size so that it is slightly narrower than the column.
\caption{\small Sensitivity to FFN expansion ratio on Beauty and Toys.}
\label{fig:Sensitivity-ffn}
\end{figure}

\noindent{\textbf{FFN Ratio}}\quad  As shown in Fig.~\ref{fig:Sensitivity-ffn}, we adjusted the FFN ratio to \{1,2,4,8\}. NDCG@20 reached its peak at \{2,4\}, while HR@20 reached its peak at 8. The FFN ratio controls the capacity of nonlinear mapping in each convolutional block. Setting it too low limits the expressive power of the model, while setting it too high may lead to overfitting or gradient instability and affect inference speed. NextCovRec focuses on the balance between inference efficiency and performance, requiring a reasonable ratio (e.g., FFN ratio = \{2,4\}).

\subsection{Efficiency and Complexity Analysis\ (RQ4)}
\begin{figure}[htbp]
\centering
\includegraphics[width=0.48\textwidth]{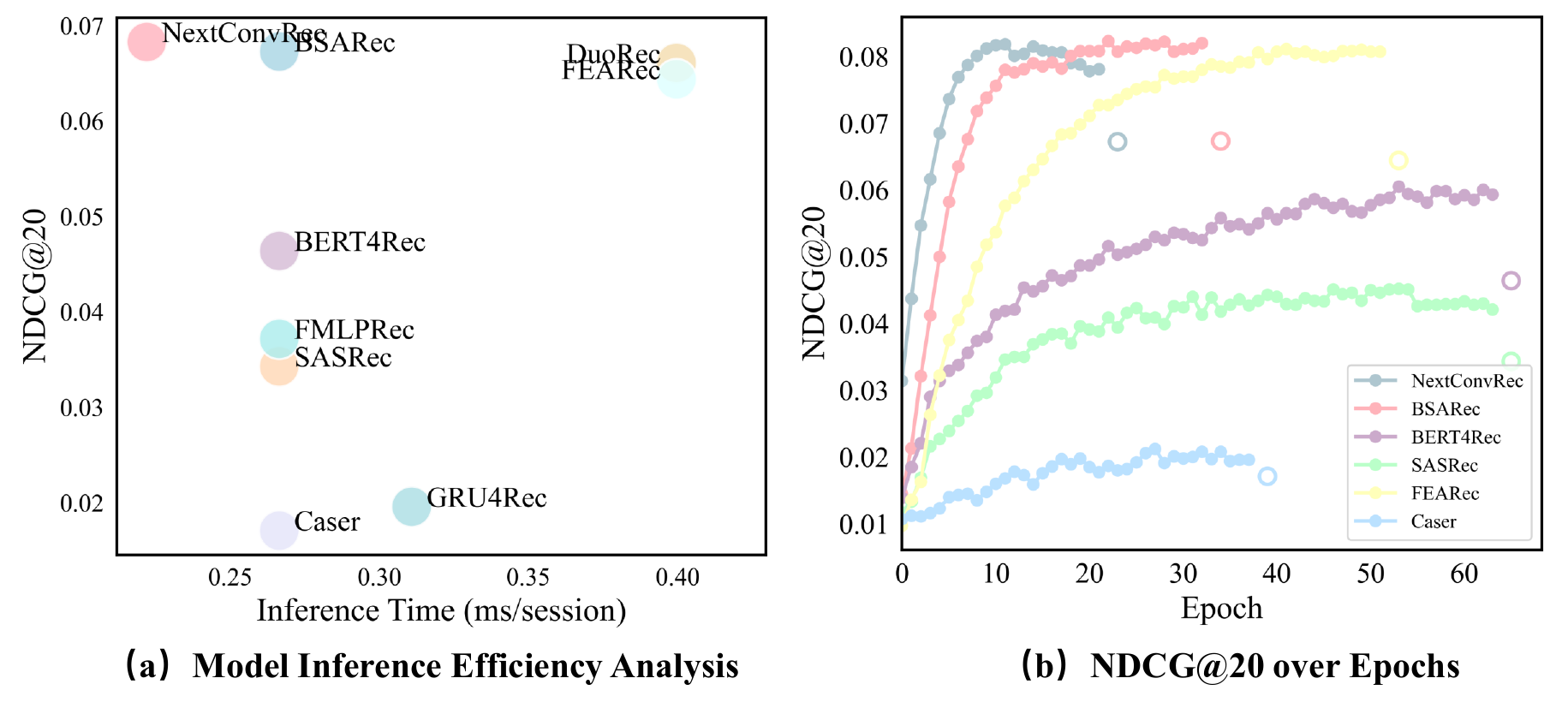} % Reduce the figure size so that it is slightly narrower than the column.
\caption{\small Analysis of inference efficiency and convergence speed of the model on the Beauty dataset. The metric for measuring inference speed is the number of sessions that can be inferred per millisecond, and the metric for measuring performance is NDCG@20.}
\label{fig:efficiency}
\end{figure}

\noindent{\textbf{Inference Time}}\quad Fig.~\ref{fig:efficiency}a presents a joint analysis of the average inference speed and performance metric NDCG@20 for the model. NextConvRec achieves an average inference time of just 0.22 ms per session while attaining an NDCG@20 of 0.0683, demonstrating optimal inference efficiency while maintaining high accuracy. The state-of-the-art Transformer-based model BSARec achieves performance comparable to NextCovRec, but its inference efficiency improves by approximately 16.7\%, due to the efficient modern convolutional architecture of NextCovRec. It is worth noting that while models like DuoRec achieve performance improvements through contrastive learning, their inference overhead is correspondingly high, which can impose significant computational burdens in industrial applications.

\noindent{\textbf{Convergence Speed}}\quad To measure the convergence speed of the model, Fig.~\ref{fig:efficiency}b shows the curves of NDCG@20 of each model as the epoch changes during the training phase. NextConvRec can quickly reach its performance limit in fewer training epochs (approximately 22 epochs on average), which is significantly faster than other models, indicating that its structural design has good training stability and optimization efficiency.

% \noindent{\textbf{Complexity of Inference}}\quad 

\section{Conclusion and Future Work}
In this paper, we explore the potential of convolutional structures in SBR tasks. Although convolutional models offer efficiency advantages, their global modeling capabilities still lag behind those of Transformer models. To address this, we combine graph structures with position-aware mechanisms for feature preprocessing and design a backbone module based on modern convolutional techniques to expand the effective receptive field and enhance global modeling capabilities. This framework achieves performance comparable to Transformer models on 4 datasets while maintaining efficiency, significantly improving inference speed and balancing performance and efficiency. In the future, we will explore lighter, more efficient convolutional variants to further reduce computational complexity and adapt to large-scale session data.

\bibliography{aaai2026}


\end{document}